\documentclass[twocolumn]{aastex61}

\newcommand{\rhoFRB}{\rho_\mathrm{FRB}}
\newcommand{\DMex}{{\rm DM_{EX}}}
\newcommand{\DMigm}{{\rm DM_{IGM}}}

\newcommand{\Lapp}{L_{\nu,{\rm app}}}
\newcommand{\Last}{L_{\nu,{\rm eff}}}
\newcommand{\Lint}{L_{\nu,{\rm int}}}

\newcommand{\Sapp}{S_{\nu,{\rm app}}}
\newcommand{\Fapp}{F_{\nu,{\rm app}}}
\newcommand{\Sth}{S_{\nu,{\rm th}}}
\newcommand{\Lo}{L_{\nu, 0}}
\newcommand{\Pks}{P_{\rm KS}}

\received{\today}
\revised{\today}
\accepted{\today}
\submitjournal{ApJ}

\shorttitle{Recipes for the statistical properties of FRBs}
\shortauthors{Niino}

\begin{document}

\title{Fast radio bursts' recipes for the distributions of dispersion measures, flux densities, and fluences}

\correspondingauthor{Yuu Niino}
\email{yuu.niino@nao.ac.jp}

\author[0000-0002-0786-7307]{Yuu Niino}
\affil{National Astronomical Observatory of Japan, 2-21-1 Osawa, Mitaka, Tokyo, Japan}

\begin{abstract}

We investigate how the statistical properties of dispersion measure (DM)
and apparent flux density/fluence of (non-repeating) fast radio bursts (FRBs) 
are determined by unknown cosmic rate density history [$\rhoFRB (z)$]
and luminosity function (LF) of the transient events. 
We predict the distributions of DMs, flux densities, and fluences of FRBs 
taking account of the variation of the receiver efficiency within its beam, 
using analytical models of $\rhoFRB (z)$ and LF. 
Comparing the predictions with the observations, 
we show that the cumulative distribution of apparent fluences 
suggests that FRBs originate at cosmological distances 
and $\rhoFRB$ increases with redshift resembling 
cosmic star formation history (CSFH). 
We also show that a LF model with a bright-end cutoff 
at log$_{10}L_\nu$ [erg s$^{-1}$Hz$^{-1}$] $\sim$ 34 
are favored to reproduce the observed DM distribution if $\rhoFRB (z)\propto$ CSFH, 
although the statistical significance of the constraints 
obtained with the current size of the observed sample is not high. 
Finally, we find that the correlation between DM and flux density 
of FRBs is potentially a powerful tool to distinguish 
whether FRBs are at cosmological distances 
or in the local universe more robustly with future observations. 

\end{abstract}

\keywords{radio continuum: general --- intergalactic medium --- ISM: general --- methods: statistical}

\section{Introduction}
\label{sec:intro}

Fast Radio Bursts (FRBs) are transient events 
observed in $\sim$ 1 GHz radio bands 
with typical durations of several milliseconds 
\citep[e.g.,][]{Lorimer:2007a, Keane:2012a, Thornton:2013a}. 
Their large dispersion measures (DMs), which indicate 
the column density of free electrons along the sightlines, 
suggest that they are extragalactic objects. 
If FRB DMs arise from the intergalactic medium (IGM), 
FRBs may provide us with an unprecedented opportunity to detect the IGM directly. 

However, the origin of FRBs is not known yet. 
Although various theoretical models have been proposed 
\citep[e.g.,][]{Totani:2013a, Kashiyama:2013a, 
Popov:2013a, Falcke:2014a, Cordes:2016b, Zhang:2017a}, 
observational evidence that confirms or rejects those models is hardly obtained. 
The currently available localization precision of FRBs are typically several arcmin, 
which is too large to identify their host galaxies, 
and FRB distance measurements 
which are independent of DM are also missing. 

The only exception is FRB~121102, the repeating FRB, 
for which the host galaxy is identified and its redshift is known 
thanks to its repeatability \citep{Chatterjee:2017a, Tendulkar:2017a}. 
However, the other FRBs do not show any repeatability, 
and hence FRB~121102 can be a different kind of phenomenon 
from the other FRBs \citep{Palaniswamy:2017a}, 
although it is also pointed out that FRB~110220 and FRB~140514 
might be repetitions of a same source \citep{Piro:2017a}. 
Hereafter, FRB means non-repeating FRB, unless stated otherwise. 

Redshift distribution of a population of transient events 
is an important clue to understand the nature of the transients, 
which reflects their luminosity function and comoving rate density at each redshift. 
The cosmic FRB rate density [$\rhoFRB (z)$] 
would be proportional to the cosmic star formation history (CSFH) 
if FRBs are related with young stellar population 
(e.g., core-collapse supernovae, young neutron stars), 
while it would follow the cosmic stellar mass density (CSMD) 
if FRBs arise from older stars. 

Although we can not measure redshift of an FRB in most cases, 
distance to an FRB can be estimated via its DM. 
The excess of the DM over the Milky Way contribution 
in the direction ($\DMex$) can be interpreted as the distance to the source
under the assumption that the major part of the observed $\DMex$ 
arise from the IGM \citep[e.g.,][]{Ioka:2003a, Inoue:2004a}.  
Previous studies have shown that the $\DMex$ distribution 
of the observed FRBs is consistent with the expectations 
if FRBs are distributed over cosmological distance 
\citep[e.g.,][]{Dolag:2015a, Katz:2016a, Caleb:2016a, Cao:2017a}. 
However, $\DMex$ does not necessarily arise only from the IGM, 
because a part of $\DMex$ can be attributed to ionized gas in the vicinity of FRBs. 

The cumulative distribution of FRB flux densities/fluences 
(so called log$N$-log$S$ distribution)
also serves as a clue to understand the distance distribution of FRBs, 
because the distribution follows a power-law with the index of -1.5 
when the sources are uniformly distributed in an Euclidian space 
while the distribution may vary when the sources 
are at cosmological distances due to the cosmic expansion 
and cosmological evolution of the source number density 
\citep{Katz:2016a, Vedantham:2016a, Caleb:2016a, Oppermann:2016a, Li:2017a, Macquart:2018a}. 

In this study, we investigate how the interplay between 
unknown cosmic rate density history and luminosity function of FRBs 
determines the statistical properties of the observable quantities, 
i.e., $\DMex$ and apparent flux density/fluence, taking account 
of the variation of the receiver efficiency within its beam. 
We discuss what constraint the current observations put on the nature of FRBs, 
and how can we distinguish whether FRBs are at cosmological distances 
or in the local universe with future observations. 

In \S\ref{sec:FRBrate} and \S\ref{sec:FRBlum}, we describe 
our models of cosmic FRB rate history and FRB luminosity function (LF), respectively. 
We discuss constraints on the cosmic rate history and the LF of FRBs 
obtained from the observed $\DMex$ distribution under the assumption 
that FRBs originate at cosmological distances in \S\ref{sec:fitting}. 
In \S\ref{sec:cosloc}, we discuss the log$N$-log$S$ distribution 
and the correlation between $\DMex$ and apparent flux density of FRBs, 
comparing the predictions of the cosmological and local FRB models. 
In \S\ref{sec:discussion}, we discuss a couple of uncertainties 
that may potentially affect our results. 
Our conclusions are summarized in \S\ref{sec:conclusion}. 
Throughout this paper, we assume the fiducial cosmology 
with $\Omega_{\Lambda}=0.7$, $\Omega_{m}=0.3$, 
and $H_0=$ 70 km s$^{-1}$ Mpc$^{-1}$. 

\section{Cosmic FRB rate history and $\DMigm$ distribution}
\label{sec:FRBrate}

\begin{figure}[t!]
\plotone{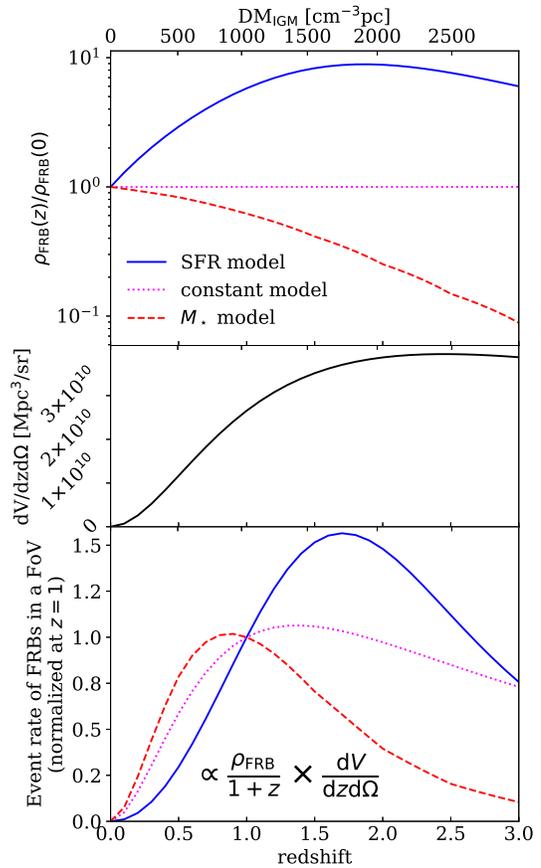}
\caption{
Top panel: $\rhoFRB$ models 
(occurrence rate of FRBs per comoving volume) normalized at $z=0$. 
Middle panel: observed comoving volume per redshift 
per steradian with the assumed cosmology ($dV/dzd\Omega$). 
Bottom panel: occurrence rate of FRBs per redshift 
per steradian in the observer frame, 
which is proportional to $\rhoFRB(z)/(1+z) \times dV/dzd\Omega$. 
We note that the FRB rates shown in this figure 
represents all FRBs regardless of their detectability. 
$\DMigm$ that corresponds to each redshift 
is indicated in the upper horizontal axis (see equation~\ref{eq:DMigm}). 
\label{fig:zdist}
}
\end{figure}

We consider three models of $\rhoFRB$ in this study
(the top panel of figure~\ref{fig:zdist}). 
One is proportional to CSFH (SFR model), 
another is constant throughout cosmic time (constant model), 
and the other is proportional to CSMD ($M_\star$ model). 
We use the formulations of CSFH and CSMD by \citet{Madau:2014a}. 

The redshift distribution of FRBs that occur 
in a unit area on the sky within a certain time period 
in the observer frame can be expressed as 
\begin{equation}
\frac{dN(z)}{dzd\Omega} = \frac{\rhoFRB(z)}{1+z} \times \frac{dV}{dzd\Omega}, 
\end{equation}
where $dV/dzd\Omega$ is comoving volume 
per redshift per observed area (the middle panel of figure~\ref{fig:zdist}). 

DM that arise from the IGM can be expressed as: 
\begin{equation}
\DMigm = c\int_0^z|\frac{dt}{dz^\prime}|
\frac{n_{e,\mathrm{IGM}}(z^\prime)}{1+z^\prime}dz^\prime. 
\label{eq:DMigm}
\end{equation}
where $n_{e,\mathrm{IGM}}$ is the electron density in the IGM. 
Here we assume that the IGM is uniform at each redshift 
with the comoving density $\rho_{\rm crit}\Omega_b$, 
composed of 75\% H and 25\% He, and fully ionized 
throughout the redshift range we consider. 
Under these assumptions, the IGM electron density can be written as: 
\begin{equation}
n_{e,\mathrm{IGM}}(z)=\frac{7}{8}\frac{\rho_{\rm crit}\Omega_b}{m_p}(1+z)^3. 
\end{equation}
The upper horizontal axis of figure~\ref{fig:zdist} 
indicates $\DMigm$ that corresponds to $z$ in the lower axis 
(naively $\DMigm \sim 1000z$ cm$^{-3}$pc in this redshift range). 

In the above expression, it is assumed 
that the dominant fraction of baryons in the universe is in the IGM, 
which is true when we consider diffuse ionized gas associated 
with dark matter halos as part of the IGM \citep[e.g.,][]{Fukugita:2004a}. 
If a significant part of the IGM is associated dark matter halos, 
the IGM might be inhomogeneous in reality, 
and the inhomogeneity might affect the $\DMigm$ distribution of FRBs. 
We discuss the effect of the IGM inhomogeneity on our results in \S\ref{ssec:IGMvar}

The predicted redshift distributions with the three $\rhoFRB(z)$ models 
are shown in the bottom panel of figure~\ref{fig:zdist}. 
The redshift distributions with the different $\rhoFRB(z)$ models 
are similar with each other at $z \lesssim 1$ 
where majority of the currently known FRBs reside, 
while the redshift distributions are dramatically different at $z > 1$, 
as previously shown by \citet{Dolag:2015a} using cosmological simulations. 
We note that detectability of FRB events are not considered here 
and the redshift distributions may include FRBs that are too faint to be detected. 
We discuss fraction of detectable FRBs at each redshift in \S\ref{sec:FRBlum}. 

\section{FRB luminosity and receiver efficiency}
\label{sec:FRBlum}

\subsection{Receiver efficiency variation within a beam} 
\label{ssec:beam}

\begin{figure}
\plotone{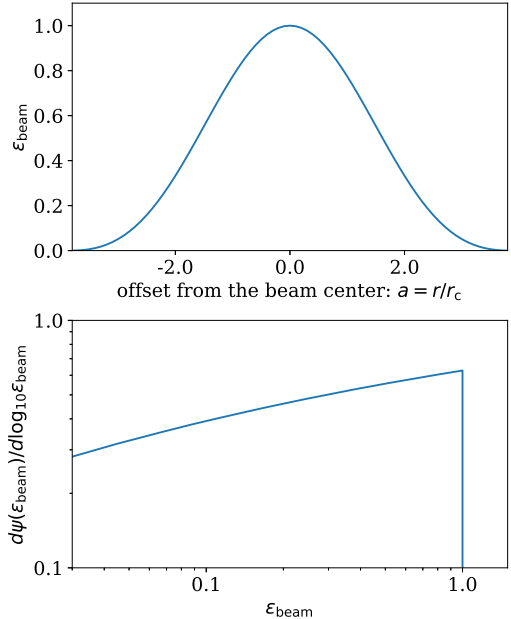}
\caption{
Top panel: the Airy disk model of a radio receiver beam efficiency 
as a function of the offset from the beam center. 
Bottom panel: the efficiency PDF within an Airy disc beam. 
\label{fig:beam}
}
\end{figure}

Observed radio flux density of an FRB at the peak of its light curve ($\Sapp$) 
depends not solely on its luminosity and distance, 
but also on the unknown position of the FRB within the receiver beam, 
because efficiency of a radio receiver largely varies within its beam. 
We assume beam efficiency pattern of a radio receiver 
under consideration is represented by an Airy disc 
\begin{equation}
\epsilon_{\rm beam}(a) = [\frac{2J_1(a)}{a}]^2,
\label{eq:epsilon}
\end{equation}
where the efficiency at the beam center is unity, 
$J_1$ is the first order Bessel function of the first kind, 
and $a = r/r_{\rm c}$ is the offset from the beam center 
normalized by the beam characteristic radius (the top panel of figure~\ref{fig:beam}). 
The efficiency is 50\% at $a = 1.62$ and drops to zero at $a = 3.83\ (\equiv a_{\rm out})$. 
For the Parkes multi-beam receiver \citep{Staveley-Smith:1996a} whose 
full width at half maximum (FWHM) is 14.4 arcmin, $r_{\rm c}$ is 4.4 arcmin. 
We do not consider sidelobe detections ($|a| > a_{\rm out}$). 

The probability distribution function (PDF) of $\epsilon_{\rm beam}$ 
within a beam ($|a| \leq a_{\rm out}$) can be written as
\begin{equation}
\frac{d\psi(\epsilon_{\rm beam})}{d{\rm log_{10}}\epsilon_{\rm beam}}
 = \frac{\rm 2ln10}{a_{\rm out}^2}\epsilon_{\rm beam}a(\epsilon_{\rm beam})
\left|\frac{da}{d\epsilon_{\rm beam}}\right|
\label{eq:PDFeff}
\end{equation}
where $a(\epsilon_{\rm beam})$ is the inverse function 
of equation~(\ref{eq:epsilon}) in the range $a > 0$. 
In the bottom panel of figure~\ref{fig:beam}, 
we show the PDF defined by equation~(\ref{eq:PDFeff}). 
We note that the PDF is not dependent of the choice of $r_{\rm c}$, 
and hence applicable to any radio telescope 
whose efficiency profile can be represented by an Airy disc. 

\subsection{Propagation effects and $K$-correction} 
\label{ssec:Lmod}

Flux density of a FRB is also affected by its propagation medium. 
While scattering may suppress FRB flux density by pulse broadening, 
scintillation and plasma lensing may also enhance FRB flux density 
\citep[e.g.,][]{Hassall:2013a, Cordes:2016a, Cordes:2017a}. 
Currently it is difficult to separate the intrinsic LF of FRBs 
from the PDF of the propagation effects. 
In this study, we consider effective luminosity ($\Last$) 
which includes the propagation effects ($\epsilon_{\rm prop}$)
rather than intrinsic luminosity ($\Lint$) of an FRB. 
We also consider apparent luminosity ($\Lapp$) which includes $\epsilon_{\rm beam}$ 
in addition to $\epsilon_{\rm prop}$ and can be directly related to $\Sapp$. 

$K$-correction is also an important effect when we consider 
observed flux densities of objects at cosmological distances. 
We express the $K$-correction factor as: 
\begin{equation}
\kappa_\nu(z) = \frac{L_\nu(\nu_{\rm rest})}{L_\nu(\nu_{\rm obs})}, 
\label{eq:Kdef}
\end{equation}
where $\nu_{\rm obs}$ is the observing frequency and $\nu_{\rm rest} = (1+z)\nu_{\rm obs}$. 
In the case of the Parkes multi-beam receiver, $\nu_{\rm obs} = 1.4$ GHz. 
The functional form of $\kappa_\nu(z)$ is determined 
by spectra of FRBs which is not known yet. 
Here we asume $\kappa_\nu(z) = 1$ (constant), 
and discuss how our results are affected by $K$-correction in \S\ref{ssec:Kcorr}. 

In summary, 
\begin{eqnarray}
\Sapp(\nu_{\rm obs}) &=& \frac{1+z}{4 \pi d_L(z)^2}\kappa_\nu(z) \Lapp(\nu_{\rm obs}),\ {\rm and} \\
\Lapp(\nu_{\rm obs}) &=& \epsilon_{\rm beam}\Last(\nu_{\rm obs})  \\
          &=& \epsilon_{\rm beam}\epsilon_{\rm prop}\Lint(\nu_{\rm obs}),   
\end{eqnarray}
where $d_L(z)$ is luminosity distance at redshift $z$. 

\subsection{FRB luminosity function} 
\label{ssec:LF}
 
We examine the following three $\Last$ distribution function models
to demonstrate how difference of FRB LF
affects the observable properties of FRBs. 
\begin{itemize}
\item LF1: FRBs are standard candles with $\Last = \Lo$. 
\item LF2: $\Last$ follows a power-law distribution, 
$d\phi(\Last)/d\Last \propto \Last^{-2}$, with a faint-end cutoff at $\Lo$. 
\item LF3: $\Last$ follows a distribution function 
with the faint-end power-law index $-1$ 
down to log$_{10}\Last$ [erg s$^{-1}$Hz$^{-1}$] = 30.0, 
and exponential cutoff in the bright-end, $\Last \gtrsim \Lo$, 
i.e., $d\phi(\Last)/d\Last \propto x^{-1}exp(-x)$, where $x = \Last/\Lo$. 
\end{itemize}

\begin{figure*}
\plotone{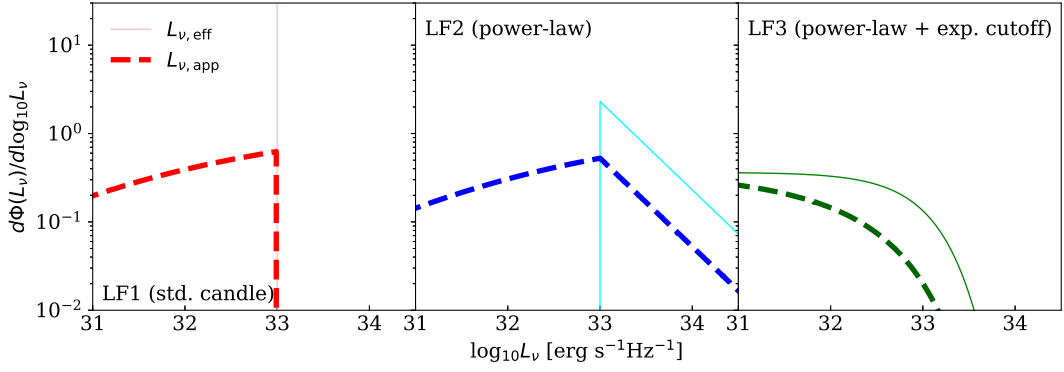}
\caption{
FRB luminosity function models considered in this study. 
The left, middle, and right panels show the PDFs of $\Last$ (thin solid) 
and $\Lapp$ (thick dashed) for LF1 (standard candles), 
LF2 (power-law), and LF3 (power-law + exponential cutoff), respectively. 
The PDFs with log$_{10}\Lo$ [erg s$^{-1}$Hz$^{-1}$] = 33 are shown for each model. 
\label{fig:LF}
}
\end{figure*}

The three $\Last$ PDFs and the corresponding $\Lapp$ PDFs are shown in figure~\ref{fig:LF}. 
The $\Lapp$ PDFs are obtained by convoluting the $\Last$ PDFs 
with the $\epsilon_{\rm beam}$ PDF (equation~\ref{eq:PDFeff}). 
The faint-end cutoff of LF3 is adopted so that the integral of the LF is finite. 
The cutoff luminosity can be observed at redshifts 
only up to $z \sim 0.01$ with the Parkes multi-beam receiver, 
and hence it is faint enough not to affect our result. 

Although the shape of the bright-end 
of the $\Lapp$ PDFs resembles that of the $\Last$ PDFs, 
the faint-end of the $\Lapp$ PDFs is smeared out by the $\epsilon_{\rm beam}$ variation. 
Hence it will be difficult to constrain the faint-end of the $\Last$ PDFs 
from the currently observable properties of FRBs. 
Although the actual shape of the FRB LF is hardly known, 
we consider the three LF models described above 
can represent a wide variety of LFs due to this smearing. 
We note that the PDF of $\epsilon_{\rm prop}$, and hence the $\Last$ PDF, 
may depend on galactic latitude and longitude of observation fields, 
if the propagation effects in the Milky Way significantly affect observed flux densities. 
However, we assume that all FRBs under consideration 
arise from a single $\Last$ PDF and consider the PDF 
as the average of those in all observation fields. 

\subsection{Detection of an FRB} 
\label{ssec:detect}

To constrain the FRB models, we use the sample of FRBs 
detected by the Parkes multi-beam receiver 
before the end of 2017 November (21 FRBs between 010125 and 160102). 
The properties of the observed FRBs are taken 
from the FRBCAT\footnote{http://frbcat.org} \citep{Petroff:2016a}. 
Although the values in the FRBCAT are derived separately by individual authors, 
\citet{Petroff:2016a} have reanalyzed some of the data in a uniform manner, 
and we use the values derived by the reanalysis when available. 
We note that FRBs discovered by different telescopes should not be treated together 
in a statistical study of the DM distribution because the $\DMigm$ distribution 
of a sample of FRBs would depend on the detection limit of the observations. 

We compute the fraction of detectable FRBs 
at each redshift using the $\Lapp$ distribution functions. 
For simplicity, we consider an FRB is detected when 
the apparent flux density exceeds a threshold, $\Sapp \geq \Sth$. 
To compare our model predictions with the Parkes detected FRB sample, 
we assume the threshold flux density $\Sth = 0.4$ Jy 
which is comparable to the faintest FRBs in the Parkes sample. 

\begin{figure*}
\plotone{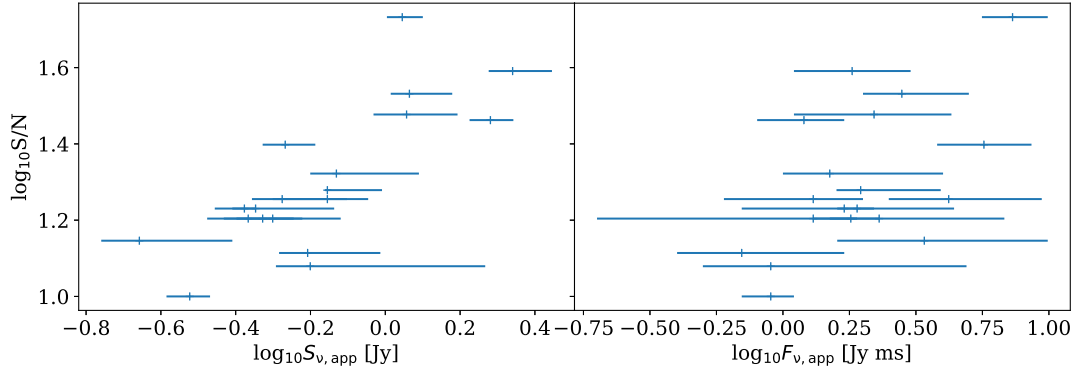}
\caption{
Left panel: the correlation between $\Sapp$ and S/N in the Parkes sample. 
Right panel: same as the left panel but between $\Fapp$ and S/N. 
The two peculiarly bright events, FRB~010724 and 150807, are excluded. 
\label{fig:F2SN}
}
\end{figure*}

\begin{figure*}
\plotone{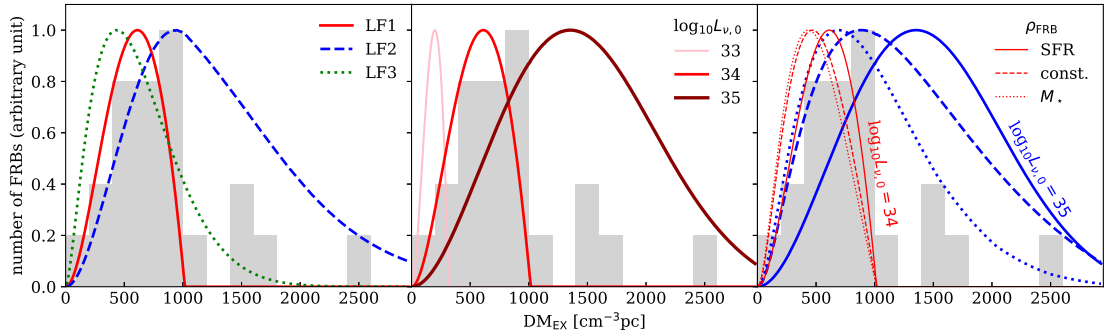}
\caption{
The $\DMigm$ distributions of detectable FRBs predicted with our models. 
LF1 with log$_{10}\Lo$ [erg s$^{-1}$Hz$^{-1}$] = 34, 
and the SFR model of $\rhoFRB$ are used, unless otherwise stated. 
The DM distribution of the Parkes sample is shown together (gray histogram). 
Left panel: the predicted $\DMigm$ distributions with the three different LF models. 
Middle panel: the $\DMigm$ distributions with different $\Lo$. 
Right panel: the $\DMigm$ distributions with the three different $\rhoFRB$ models. 
Thin and thick lines indicate the distributions 
with log$_{10}\Lo$ [erg s$^{-1}$Hz$^{-1}$] = 34 and 35, respectively. 
\label{fig:distDMigm}
}
\end{figure*}

Although it is pointed out that detectability of an FRB depends 
not only on its flux but also on the pulse width 
\citep[and hence the fluence,][]{Keane:2015a}, 
the Parkes sample shows that $\Sapp$ is a better proxy 
for signal-to-noise ratio (S/N) than apparent fluence 
[observed fluence including $\epsilon_{\rm beam}$ ($\Fapp$), see figure~\ref{fig:F2SN}]. 
When the saturated event FRB~010724 \citep{Lorimer:2007a} 
and the extremely bright outlier event 
FRB~150807 \citep{Ravi:2016a} are excluded from the sample, 
the correlation coefficient between log$_{10}\Sapp$ and log$_{10}$S/N is 0.79 
(0.58 between log$_{10}\Fapp$ and log$_{10}$S/N). 
In figure~\ref{fig:distDMigm}, we show how the predicted 
$\DMigm$ distribution of detectable FRBs depends on the assumed FRB models. 

\section{Luminosity of FRBs in the case that they originate at cosmological distances}
\label{sec:fitting}

\begin{figure*}
\plotone{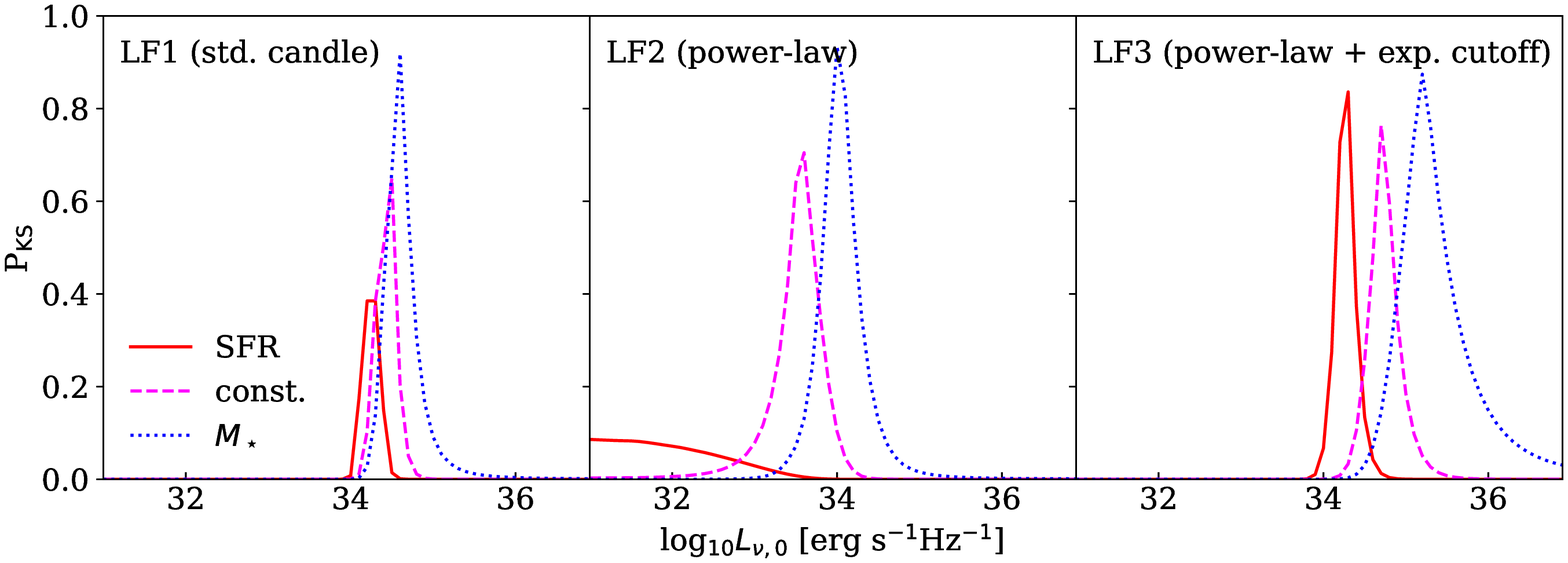}
\caption{
$\Pks$ likelihood between the model 
and observed $\DMex$ distributions as a function of $\Lo$. 
The left, middle, and right panels 
show $\Pks$ for LF1, LF2, and LF3, respectively. 
The cases with the three $\rhoFRB$ models are shown for each LF. 
\label{fig:DMfitPks}
}
\end{figure*}

\begin{figure*}
\plotone{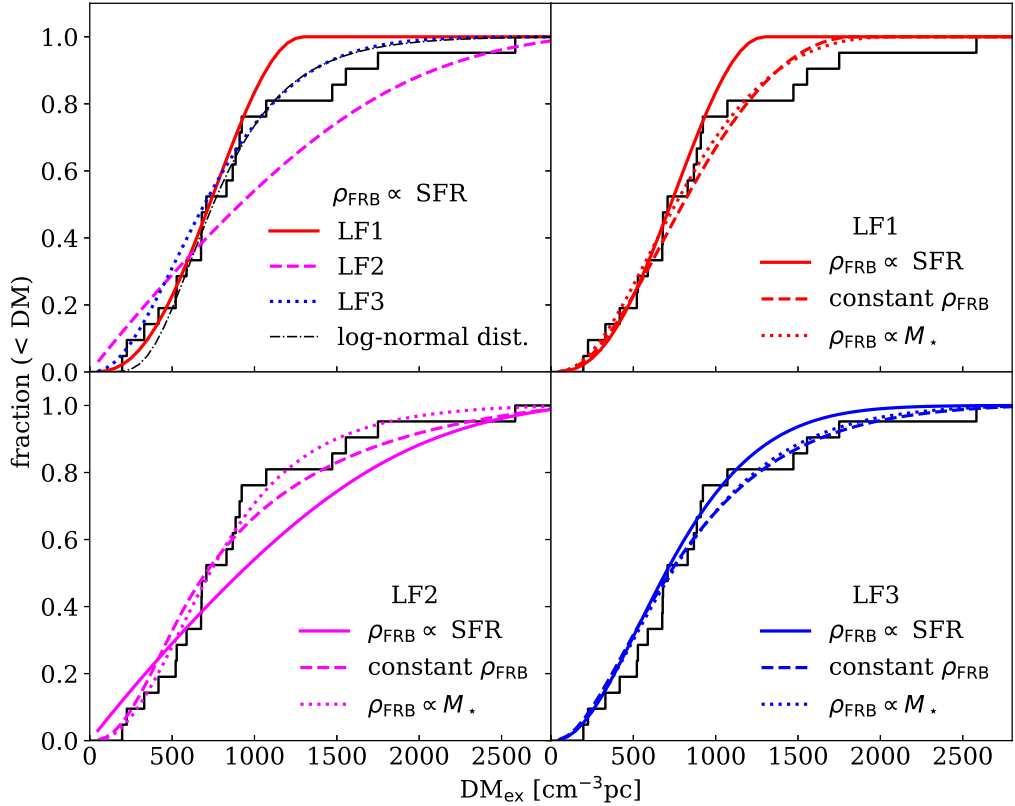}
\caption{
The cumulative $\DMex$ distribution of the Parkes FRB sample (histogram), 
and the best fit model distributions to the observations.
Top left panel: the best fit models with LF1, LF2, and LF3 
[solid (red), dashed (magenta), and dotted (blue) lines, respectively], 
the SFR model of $\rhoFRB$ is assumed. 
A simple log-normal distribution 
with the median $\DMex = 750$ cm$^{-3}$pc and $\sigma = 0.2$ dex 
is plotted together [dot-dashed line (black)]. 
Top right, bottom left, and bottom right panels: 
the best fit distributions with the different $\rhoFRB$ models 
(SFR: solid line, constant: dashed line, and $M_\star$: dotted line) 
with LF1, LF2, and LF3, respectively. 
\label{fig:DMmodeldists}
}
\end{figure*}

Here we determine the characteristic luminosity density 
of FRBs ($\Lo$, see \S\ref{ssec:LF}) that reproduces 
the observed $\DMex$ distribution best for each set of LF and $\rhoFRB$ models 
assuming that FRBs originate at cosmological distances 
and the observed $\DMex$ is dominated by $\DMigm$. 
We evaluate the goodness of fit using the Kolmogorov-Smirnov (KS) test. 
Figure~\ref{fig:DMfitPks} shows the KS test probability ($\Pks$) that the observed sample 
can arise from the model distribution as a function of $\Lo$, 
and figure~\ref{fig:DMmodeldists} shows the best fit $\DMigm$ distributions. 

Although a wide variety of LF and $\rhoFRB$ models 
agree with the observed $\DMex$ distribution, 
the model with $\rhoFRB\propto$ SFR plus LF2 
does not reproduces the observations well. 
The best fit $\Lo$ for the $\rhoFRB\propto$ SFR plus LF2 model 
is log$_{10}\Lo$ [erg s$^{-1}$Hz$^{-1}$] $\leq 31$ ($\Pks$ is constant for smaller $\Lo$), 
which is smaller than the best fit values for the other models. 
This is because LF2 makes the $\DMigm$ 
distribution heavily tailed in the high DM end, 
while the observed $\DMex$ distribution 
steeply declines above $\DMex\gtrsim 1000$ cm$^{-3}$pc (the left penel of figure~\ref{fig:distDMigm}). 
The small $\Lo$ suppresses the high DM tail in the model distribution, 
and minimize the discrepancy between the model and observed distribution. 
However, it also overpredicts the number of FRBs with $\DMex \lesssim 500$ cm$^{-3}$pc
making the model distribution broader than observed. 

The discrepancy between the model 
with $\rhoFRB\propto$ SFR plus LF2 and the observations suggests 
that neither an FRB LF with an extended bright-end without cutoff, 
nor an FRB LF that is dominated by its faint-end is favorable to reproduce 
the observed narrow $\DMex$ distribution when $\rhoFRB\propto$ SFR, 
although the current FRB sample is not sufficient 
to rule out the model with high statistical significance. 
On the other hand, LF1 and LF3 reproduce 
the observations with similar $\Lo$ to each other 
(log$_{10}\Lo$ [erg s$^{-1}$Hz$^{-1}$] $\sim$ 34--35), 
indicating that the faint-end of a $\Last$ PDF 
does not significantly affect the $\DMex$ distribution 
unless the $\Last$ PDF is dominated by its faint-end as in the case 
of LF2 with log$_{10}\Lo$ [erg s$^{-1}$Hz$^{-1}$] $\lesssim 31$. 

It is also noticeable that the $\rhoFRB\propto$ SFR plus LF1 model 
produces sharp upper limit in the $\DMigm$ distribution 
which reflects the upper limit of the $\Last$ distribution 
making the agreement between the model 
and the observations poorer than those with the other models, 
although it is not rejected with a certain statistical significance. 
The decrease in the number of FRBs above $\DMigm\gtrsim 1000$ 
in the constant and $M_\star$ models (the bottom panel of figure~\ref{fig:zdist}) 
can ease the conflict between LF1/LF2 and the observations. 
In those cases, LF2 also favors log$_{10}\Lo$ [erg s$^{-1}$Hz$^{-1}$] $\sim 34$. 

\section{Are FRBs cosmological or local events?}
\label{sec:cosloc}

Although the cosmological $\DMigm$ models 
(except that with $\rhoFRB\propto$ SFR plus LF2) 
reproduce the observed $\DMex$ distribution well as previously shown 
by \citet{Dolag:2015a}, \citet{Katz:2016a}, and \citet{Caleb:2016a}, 
it should be noted that the distribution 
can also be explained by a simple log-normal distribution 
with the median $\DMex = 750$ cm$^{-3}$pc and $\sigma = 0.2$ dex 
(shown in the top left panel of figure~\ref{fig:DMmodeldists}). 
Since a log-normal distribution is one of the most commonly seen PDFs in nature, 
it can be easily produced by a population of ionized gas 
in the circum/inter-stellar medium (CSM/ISM) around FRB sources 
without significant contribution from the IGM. 
Although $\DMex$ as high as 750 cm$^{-3}$pc is not 
likely to arise from ordinary galaxy ISM, 
if FRB sources are associated with ionized gas such as supernova remnant, 
it may significantly contribute to the observed $\DMex$
\citep{Connor:2016a, Piro:2016a, Murase:2016a, Lyutikov:2016a}. 

Here, we discuss how to distinguish 
whether FRBs are at cosmological distances 
(cosmological FRB model, $\DMex$ is dominated by $\DMigm$) 
or in the local universe (local FRB model, $\DMex$ is dominated 
by CSM/ISM in the vicinity of FRB sources).  

\subsection{log$N$-log$S$ distribution}
\label{ssec:logNlogS}

When a population of light source is homogeneously distributed 
in a Euclidean space as in the case of the local FRB model, 
observed flux density and fluence of the sources (so called log$N$-log$S$ distribution) 
follow the power-law distribution $N(< S_\nu) \propto S_\nu^\alpha$ with index $\alpha = -1.5$. 
Although actual $S_\nu$ and $F_\nu$ of an FRB is not measurable 
due to the uncertain beam efficiency for each event, 
$\Sapp$ and $\Fapp$ would also follow the same power-law distribution 
when actual $S_\nu$ and $F_\nu$ follows the power-law distribution. 
Thus the observed distributions of $\Sapp$ and $\Fapp$ 
can serve as clues to distinguish whether FRBs are cosmological or local. 

The earlier studies by \citet{Vedantham:2016a}, \citet{Caleb:2016a}, and \citet{Li:2017a}
showed that the $\Fapp$ distribution is flatter than the Euclidean case ($\alpha > -1.5$). 
However, \citet{Macquart:2018a} pointed out 
that the $\Fapp$ distribution of FRBs are largely affected 
by the detection incompleteness in the faint-end, 
and the steepness of the distribution is dependent 
on the fluence limit applied in the analysis. 
The recent analyses by \citet{Macquart:2018a} and \citet{Bhandari:2017a} 
showed that the observed FRB sample indicates 
that the $\Fapp$ distribution is steeper than the Euclidean case ($\alpha < -1.5$)
above the fluence completeness limit of 2 Jy ms \citep{Keane:2015a}, 
although the current FRB sample size is not sufficient to reject the Euclidean case. 
On the other hand, \citet{Oppermann:2016a} examined 
the distribution of S/N of FRBs rather than $S_\nu$ and $F_\nu$, 
and found that the log$N$-log$S$ distribution agrees well with the Euclidean case. 

\begin{figure*}
\plotone{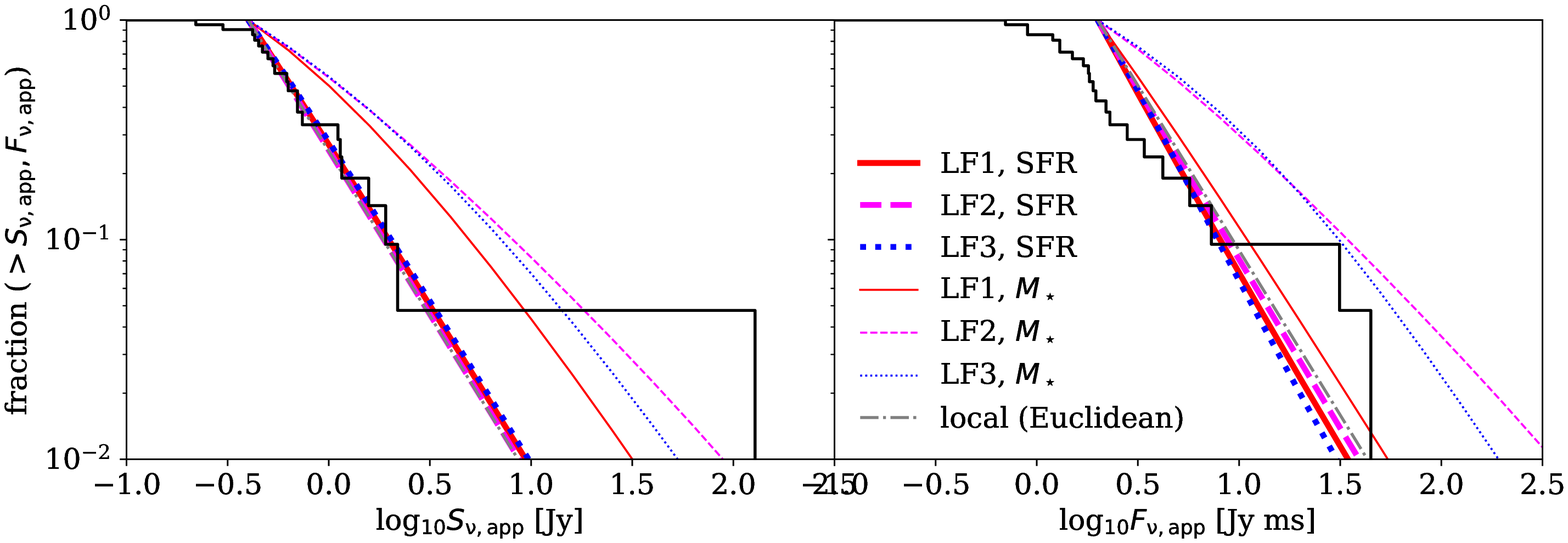}
\caption{
Left panel: the cumulative distribution of $\Sapp$ (log$N$-log$S$), 
predicted by the cosmological FRB models with 
the SFR and $M_\star$ models of $\rhoFRB$ (thick and thin lines, respectively). 
The results with the three LF models are shown 
(LF1, LF2, and LF3; solid, dashed, and dotted lines, respectively). 
The $\Sapp$ distribution of the Parkes sample is plotted together (histogram), 
and the dot-dashed line indicates the distribution 
in the Euclidean case ($\alpha = -1.5$). 
Right panel: same as the left panel but for $\Fapp$.  
\label{fig:logNlogF}
}
\end{figure*}

In the left panel of figure~\ref{fig:logNlogF}, 
we show the predicted $\Sapp$ distributions 
by the cosmological FRB models assuming 
the SFR and $M_\star$ models of $\rhoFRB$ 
together with the three LF models. 
Hereafter, the parameter $\Lo$ in the LF models is fixed to the best fit value 
determined in \S\ref{sec:fitting}, unless otherwise stated. 
The distribution functions predicted by the $M_\star$ model of $\rhoFRB$ 
are shallower than the Euclidean case regardless of the assumed LF model, 
while the distributions predicted by the SFR model 
of $\rhoFRB$ are similar to the Euclidean case. 
This is because the cosmological expansion makes 
the log$N$-log$S$ distribution shallower, 
while larger source density at larger distance 
(as in the case of the SFR model of $\rhoFRB$) makes the distribution steeper. 

The right panel of figure~\ref{fig:logNlogF} shows 
the same distribution as that in the left panel but for $\Fapp$. 
We have assumed the PDF of FRB energy follow 
the same formulations as the LF (LF1, LF2, and LF3), 
with the characteristic energy $E_0 = \Lo \times 3$ ms, 
and the fluence threshold to be 2 Jy ms 
which is the completeness limit derived by \citet{Keane:2015a} 
although many FRBs are detected below this fluence. 

The predicted $\Fapp$ distributions are steeper than the $\Sapp$ distributions 
because fluence is not affected by the cosmological expansion of time. 
As a result, the $\Fapp$ distribution functions predicted 
by the SFR model of $\rhoFRB$ is steeper than the Euclidean case ($\alpha \sim -1.8$), 
being consistent with the suggestions by the recent observations
\citep{Macquart:2018a, Bhandari:2017a}. 
Although the Euclidean case ($\alpha = -1.5$) is not fully ruled out by the current sample, 
if the steep fluence distribution is confirmed with the larger FRB sample, 
it indicates that FRBs originate at cosmological distances 
and $\rhoFRB$ is larger at higher redshift resembling CSFH 
(see \S\ref{ssec:Kcorr} for another possibility). 

The difference of $\alpha$ between the $\Fapp$ distribution 
and the $\Sapp$ distribution predicted by the cosmological FRB models 
with $\rhoFRB\propto$ SFR can also reconcile 
the different $\alpha$ suggested by \citet[S/N distribution]{Oppermann:2016a}  
and \citet[$\Fapp$ distribution]{Macquart:2018a}, 
given that $\Sapp$ correlates well with S/N. 
On the other hand, the shallow log$N$-log$S$ distributions 
predicted by the $M_\star$ model of $\rhoFRB$ 
are close to the upper limit of $\alpha$ derived by \citet{Amiri:2017a}, 
and hence can be rejected in the near future. 

In the current Parkes sample, 9 out of the 21 FRBs 
have larger $\Fapp$ than the 2 Jy ms completeness limit. 
\citet{Macquart:2018a} examined how precisely $\alpha$ 
can be determined for a variation of FRB sample sizes. 
Their results suggest that $\sim$ 50 FRBs above the fluence completeness limit 
would be necessary to distinguish $\alpha = -1.8$ (our model prediction) 
from the Euclidean case with a statistical significance of $\sim$ 95\%. 
If the fraction of FRBs above the fluence completeness limit 
in the observed sample remains unchanged, 
the total sample size required will be $\sim$ 100 FRBs. 

\subsection{Correlation between $\DMex$ and $\Sapp$}
\label{ssec:DMFcorrelation}

Unlike $\Fapp$, $\Sapp$ correlates well with S/N 
and the cumulative distribution of $\Sapp$ does not 
show significant incompleteness in its faint-end 
(figure~\ref{fig:F2SN} and \ref{fig:logNlogF}). 
Hence we might be able to utilize larger observed sample 
when we investigate $\Sapp$ rather than $\Fapp$. 
However, the cumulative distribution of $\Sapp$ 
of the cosmological FRB model is similar to 
the Euclidean case (i.e., the local FRB model) when $\rhoFRB\propto$ SFR, 
making it difficult to distinguish whether FRBs 
are cosmological or local solely with the $\Sapp$ distribution. 

One possible clue is the correlation between $\DMex$ and $\Sapp$. 
\citet{Yang:2017a} investigated the correlation 
between $\DMex$ and observed flux density to constrain 
the contribution of CSM/ISM in the vicinity of FRBs 
to $\DMex$ but without taking account of 
the $\epsilon_{\rm beam}$ variation within a receiver beam. 
Here we examine how efficiently the cosmological and local FRB models  
can be distinguished by the correlation between $\DMex$ 
and $\Sapp$, in the case $\rhoFRB\propto$ SFR. 

\begin{figure}
\plotone{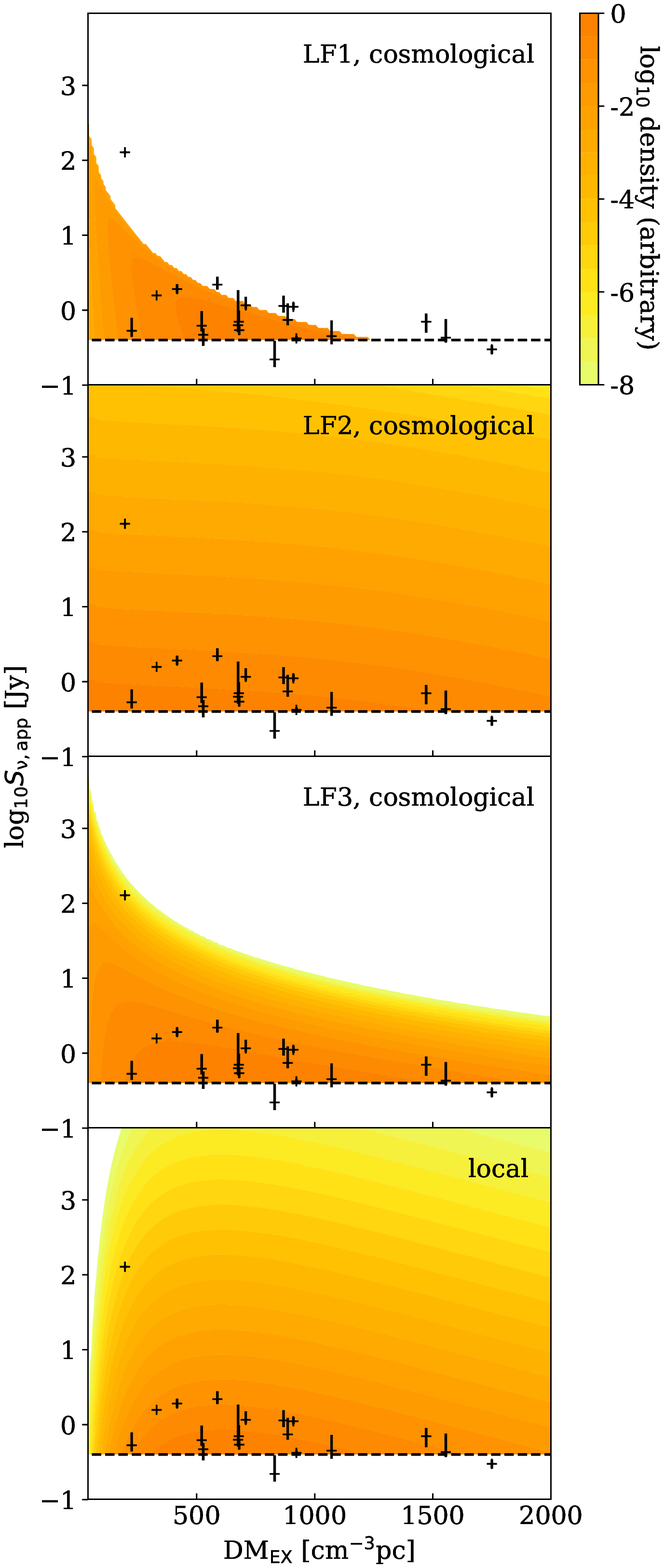}
\caption{
Distributions of FRBs on the parameter plane of $\DMex$ vs. $\Sapp$. 
The top three panels show the distributions 
of the cosmological FRBs with LF1, LF2, and LF3, respectively. 
$\rhoFRB$ is assumed to be proportional to SFR. 
The bottom panel shows the distribution of the local FRB model. 
The horizontal dashed line indicates 
the assumed detection limit in our model (0.4 Jy). 
FRBs in the Parkes sample are overplotted with crosses. 
\label{fig:DMF2D}
}
\end{figure}

We compute distribution of FRBs on the parameter plane 
of $\DMex$ vs. $\Sapp$ using the cosmological FRB models 
with LF1, LF2, and LF3 (the top three panels of figure~\ref{fig:DMF2D}). 
For the local FRB model, we assume 
that the log$N$-log$S$ distribution is the power-law with $\alpha = -1.5$, 
and $\DMex$ follows the log-normal distribution 
with the median $\DMex = 750$ cm$^{-3}$pc 
and $\sigma = 0.2$ dex (the bottom panel of figure~\ref{fig:DMF2D}). 

\begin{figure}
\plotone{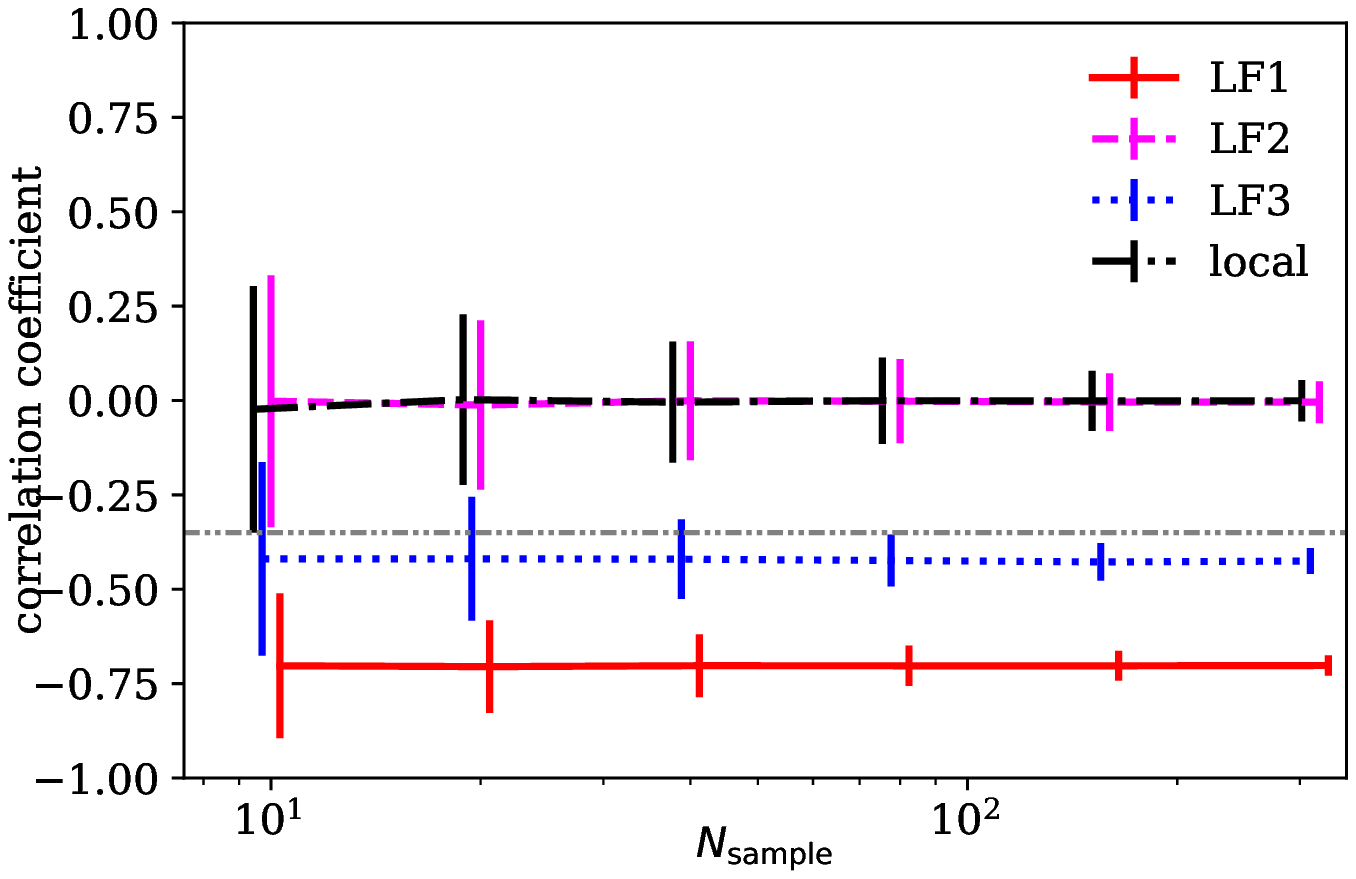}
\caption{
Mean and standard deviation 
of the correlation coefficient between $\DMex$ and $\Sapp$ 
generated by the Monte Carlo tests according 
to the distribution function shown in figure~\ref{fig:DMF2D}. 
Datapoints connected with solid, dashed, dotted lines 
show the correlation coefficient distribution with the cosmological FRB models 
with LF1, LF2, and LF3, respectively. 
The correlation coefficient distribution with the local FRB model 
is shown with datapoints connected with dot-dashed line. 
The datapoints are slightly shifted sideways for visibility. 
The horizontal double-dot-dashed line indicates 
the correlation coefficient between $\DMex$ and $\Sapp$ in the Parkes sample 
with FRB~010724 and 150807 excluded ($N_{\rm sample} = 19$). 
The random generation of mock sample 
is performed 1000 times for each $N_{\rm sample}$. 
\label{fig:CCstat}
}
\end{figure}

We then randomly generate $10^3$ sets of mock samples 
of $\DMex$ and $\Sapp$ with sample size $N_{\rm sample}$ each 
in accordance with the model distributions, 
and compute probability distribution of 
the correlation coefficient between $\DMex$ and $\Sapp$. 
In figure~\ref{fig:CCstat}, we show the mean and the standard deviation  
of the correlation coefficient distributions as functions of $N_{\rm sample}$. 

When the two peculiarly bright events, FRB~010724 and 150807, 
are excluded, the correlation coefficient between $\DMex$ and $\Sapp$ 
in the Parkes sample is -0.35 with $N_{\rm sample} = 19$. 
The correlation coefficient in the Parkes sample is already outside the standard deviation 
of the local FRB model with the current $N_{\rm sample}$. 
Although the correlation coefficient is still within the $2\sigma$ error of the local FRB model,  
it can be ruled out if the same correlation coefficient is obtained with $N_{\rm sample} = 40$. 

Among the cosmological FRB models with $\rhoFRB\propto$ SFR, 
LF3 agrees best with the observations. 
The correlation coefficient distribution with LF2 
is hardly distinguishable from that with the local FRB model, 
however LF2 is disfavored by the $\DMex$ distribution (see \S\ref{sec:fitting}). 
When the constant and $M_\star$ models of $\rhoFRB$ are assumed, 
the correlation coefficient between $\DMex$ and $\Sapp$ 
is not significantly changed with LF1 and LF3, 
while the model with LF2 shows the correlation coefficient 
of $\sim$ $-0.2$--$-0.3$ depending on the $\rhoFRB$ model. 

If FRBs with higher $\DMex$ suffer more pulse broadening, 
it is possible that $\DMex$ and $\Sapp$ correlates even 
in the local FRB model because pulse broadening may decrease $\Sapp$. 
However, we note that the pulse width of the FRBs in the Parkes sample 
is not correlated with their $\DMex$ (the correlation coefficient is -0.003). 

\section{Discussion}
\label{sec:discussion}

\subsection{Inhomogeneous IGM}
\label{ssec:IGMvar}

In the previous sections, we have assumed 
that the IGM density is spatially uniform at each redshift. 
\citet{McQuinn:2014a} computed $\DMigm$ variation of FRBs 
at a single redshift caused by the inhomogeneity of the IGM. 
Their results show that the standard deviation of $\DMigm$ is 20--30\% 
of the mean $\DMigm$ at each redshift in the range $z \sim 0.3$--1.4 
when the spatial baryon distribution from a cosmological simulation is assumed. 

\begin{figure}
\plotone{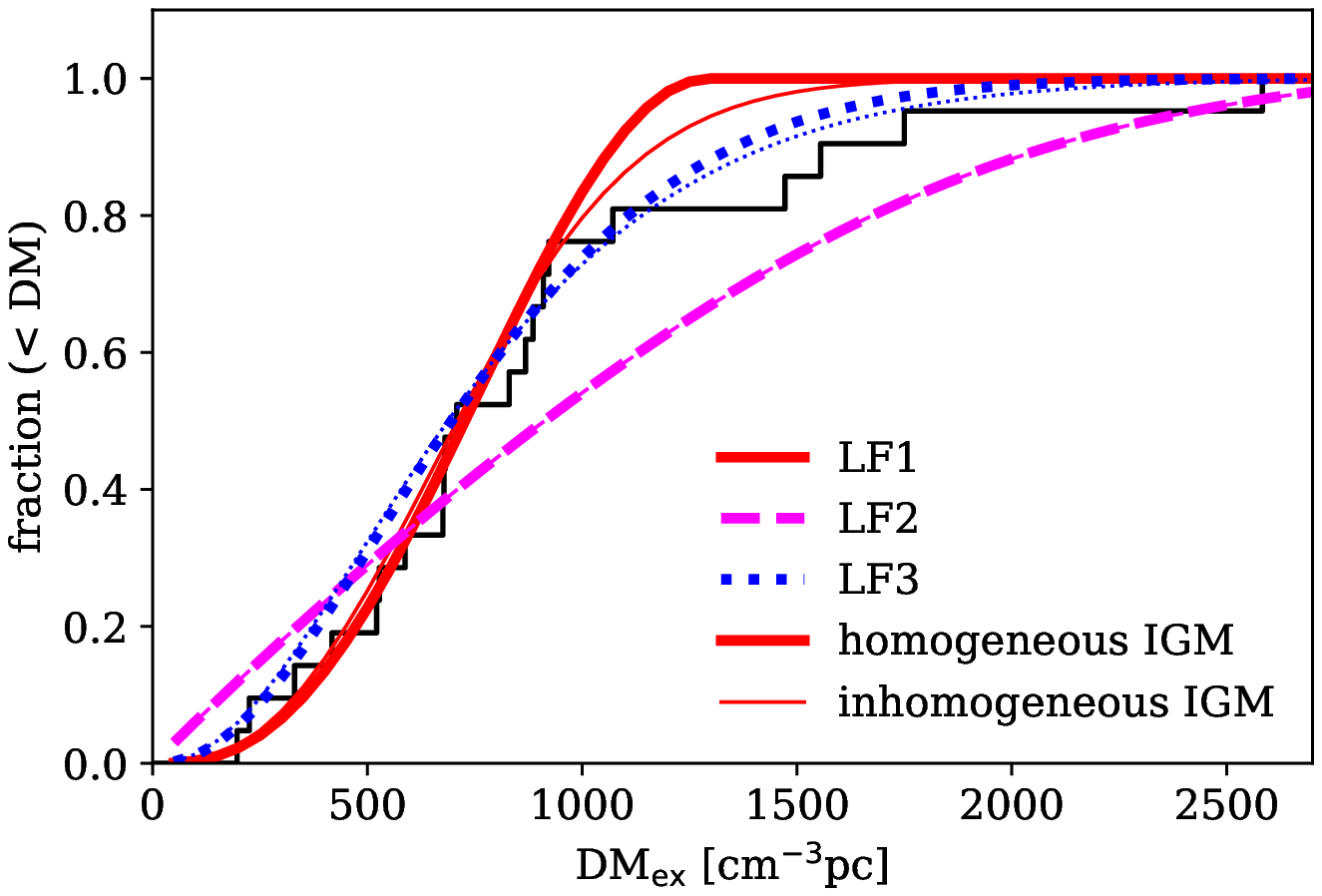}
\caption{
The cumulative $\DMex$ distributions predicted 
with (thin line) and without (thick line) the inhomogeneity of the IGM. 
The solid, dashed, and dotted lines represent 
the distributions predicted with LF1, LF2, and LF3, respectively. 
The SFR model of $\rhoFRB$ is assumed. 
The distributions without the inhomogeneity 
are identical to those in figure~\ref{fig:DMmodeldists}. 
$\DMex$ of the Parkes sample is plotted together (histogram). 
\label{fig:igmv}
}
\end{figure}

To test how such $\DMigm$ variation affect 
the overall $\DMigm$ distribution that includes FRBs at various redshits. 
We compute the $\DMigm$ distributions with the cosmological FRB models 
assuming that the probability distribution of log$_{10}\DMigm$ 
at a redshift follows a Gaussian distribution with the mean value 
determined by equation~(\ref{eq:DMigm}) and $\sigma = 0.1$ dex. 
We find that the inhomogeneity of the IGM does not 
significantly affect neither the overall $\DMigm$ distribution of FRBs, 
nor the PDF of the correlation coefficient between $\DMex$ and $\Sapp$. 
The $\DMigm$ distributions predicted with and without 
the inhomogeneity of the IGM are shown in figure~\ref{fig:igmv}.

\subsection{$K$-correction}
\label{ssec:Kcorr}

\begin{figure}
\plotone{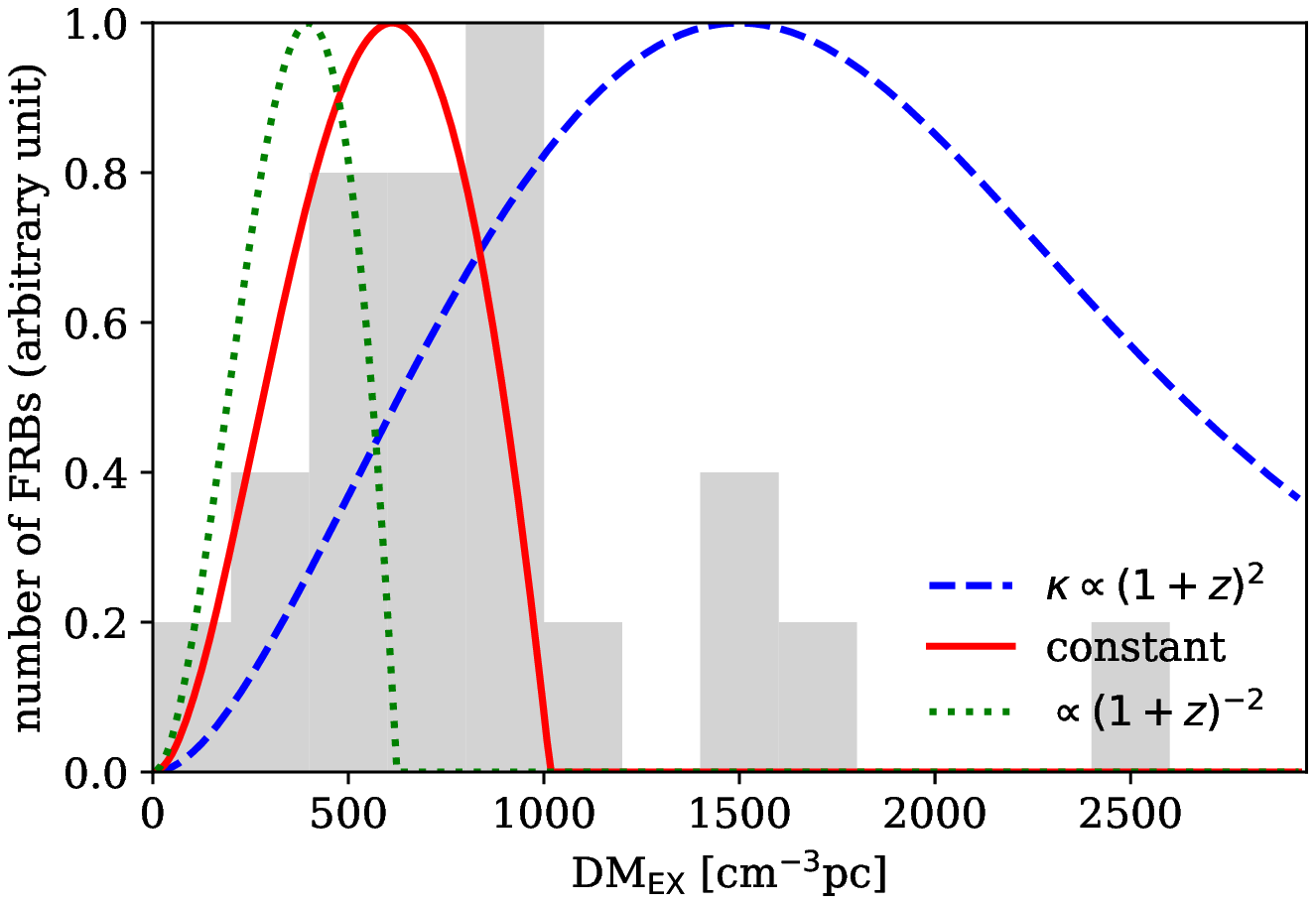}
\caption{
Same as figure~\ref{fig:distDMigm}, but with $\beta =$ -2, 0, and 2. 
LF1 with log$_{10}\Lo$ [erg s$^{-1}$Hz$^{-1}$] = 34, and the SFR model of $\rhoFRB$ are assumed. 
\label{fig:distDMigm_Kcorr}
}
\end{figure}

We have also assumed that the $K$-correction factor to be $\kappa_\nu(z) = 1$ (constant). 
The real $\kappa_\nu(z)$ is determined by spectra 
of FRBs which is not well known yet (see equation~\ref{eq:Kdef}). 
For example, when the spectrum of an FRB 
is a power-law, $\Last(\nu)\propto\nu^\beta$, 
the $K$-correction factor is $\kappa_\nu(z)=(1+z)^\beta$. 
If $\kappa_\nu(z)$ increases with redshift, 
we would detect more FRBs at higher redshifts. 
In this sense, there is a degeneracy between 
the $K$-correction (spectrum) and $\rhoFRB$. 
In figure~\ref{fig:distDMigm_Kcorr}, 
we show the $\DMigm$ distributions with $\beta =$ -2, 0, and 2, 
assuming LF1, $\rhoFRB\propto$ SFR, and log$_{10}\Lo$ [erg s$^{-1}$Hz$^{-1}$] = 34. 

\begin{figure*}
\plotone{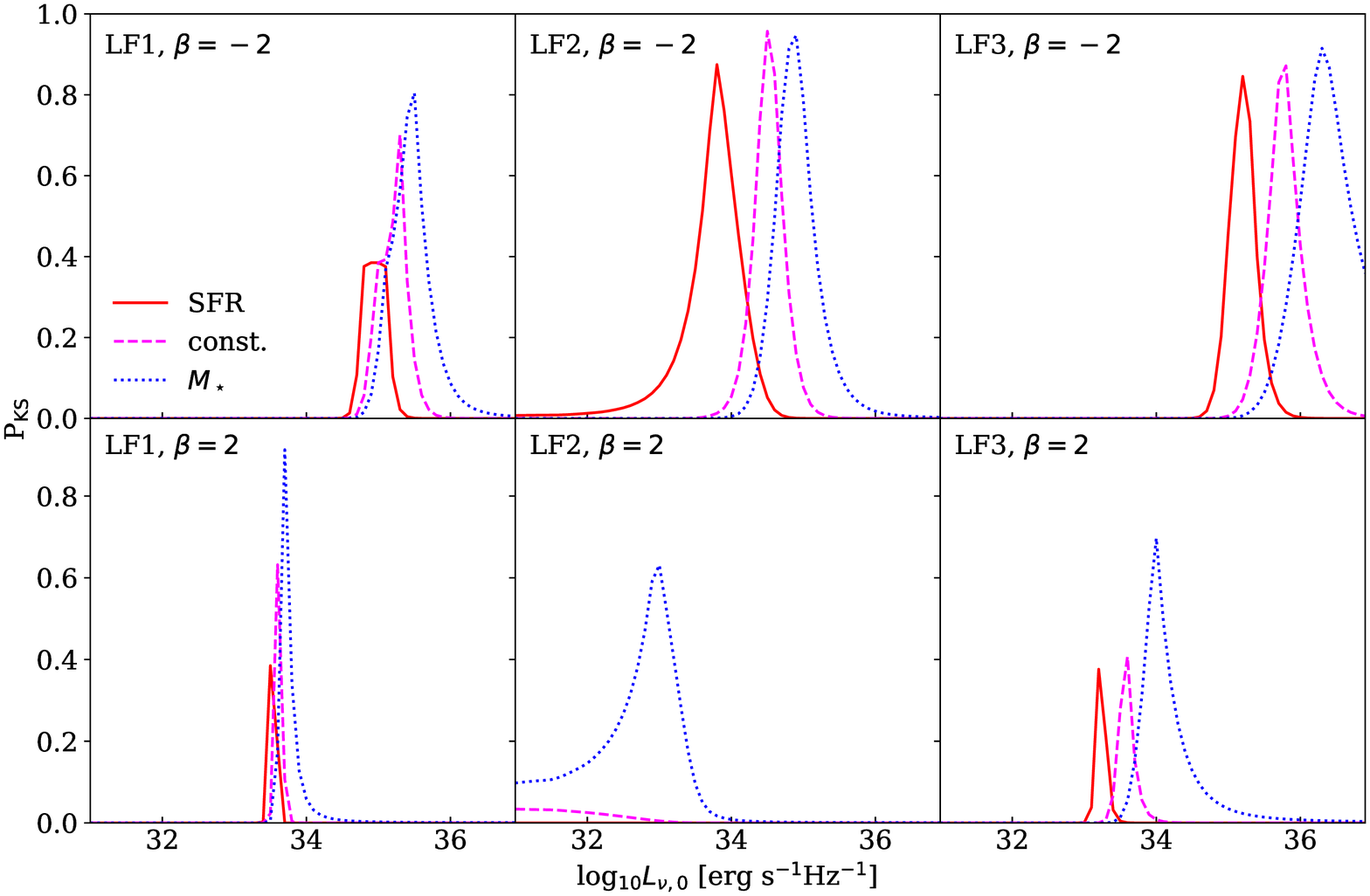}
\caption{
Same as figure~\ref{fig:DMfitPks}, but with $\beta = \pm 2$. 
\label{fig:DMfitPks_Kcorr}
}
\end{figure*}

We have determined the $\Lo$ that reproduces the observed $\DMex$ distribution 
with $\beta = \pm 2$ following the same procedure 
as in \S\ref{sec:fitting} (figure~\ref{fig:DMfitPks_Kcorr}).  
The observed $\DMex$ distribution can be reproduced 
in a wide variety of cases but with different $\Lo$. 
Once the best fit $\Lo$ for each $\beta$ is determined, 
the $K$-correction does not significantly affect 
the correlation coefficient between $\DMex$ and $\Sapp$. 
However, we note that $\beta > 0$ can also make 
the cumulative distribution of $\Fapp$ steeper 
as well as the increase of $\rhoFRB$ at high redshifts, 
due to the degeneracy between the $K$-correction and $\rhoFRB$. 
Observations with different $\nu_{\rm obs}$ 
are important to break the degeneracy. 

\section{Conclusions} 
\label{sec:conclusion}

We have computed the $\DMex$ distribution, the log$N$-log$S$ distribution, 
and the $\DMex$--$\Sapp$ correlation based on the analytic models 
of cosmic rate density history ($\rhoFRB$) and LF of FRBs. 
Comparing the model predictions with the observations, 
we have found that the cumulative distribution of apparent fluences 
suggests that FRBs are at cosmological distances 
with higher $\rhoFRB$ at higher redshifts resembling CSFH 
(or FRBs typically have very hard radio spectra 
with $L_\nu$ larger at higher frequency, i.e., $\beta > 0$), 
although the sample size of the current observations 
is not sufficient to rule out that FRBs originate in the local universe. 

If $\rhoFRB$ is proportional to SFR, 
the narrow $\DMex$ distribution of the observed FRBs 
favors an FRB LF with a bright-end cutoff 
at log$_{10}L_\nu$ [erg s$^{-1}$Hz$^{-1}$] $\sim$ 34. 
Although the constraint on the faint-end of FRB LF is rather weak, 
an FRB LF that is dominated by its faint-end is also disfavored. 
However, the statistical significance of the constraint with the current sample is still low. 

The correlation coefficient between $\DMex$ and $\Sapp$ 
is potentially a very powerful tool to distinguish 
whether FRBs are at cosmological distances or in the local universe 
more robustly with future observations, 
which may provide us with higher statistical significance 
of the constraint than the log$N$-log$S$ distribution for a given sample size.


\acknowledgments
We thank Yuan-Pei Yang and Liam D. Connor for their useful suggestions. 
Thanks are also due to the anonymous referee for his/her encouraging comments. 
This research has been supported by JSPS KAKENHI Grant Number JP17K14255. 

\bibliography{reference_list}

\end{document}